\documentclass[twocolumn,aps,prl,showpacs,amsmath,amssymb]{revtex4}
\usepackage{graphicx,dcolumn,bm}

\begin{document}

\title{Spectator Behavior in a Quantum Hall Antidot with Multiple Bound Modes}

\author{W.-R. Lee}
\author{H.-S. Sim}
\affiliation{Department of Physics, Korea Advanced Institute of Science and Technology, Daejeon 305-701, Korea}

\date{\today}

\begin{abstract}

We theoretically study Aharonov-Bohm resonances in an antidot system with multiple bound modes in the integer quantum Hall regime, taking capacitive interactions between the modes into account. We find the spectator behavior that the resonances of some modes disappear and instead are replaced by those of other modes, due to internal charge relaxation between the modes. This behavior is a possible origin of the features of previous experimental data which remain unexplained, spectator behavior in an antidot molecule and resonances in a single antidot with three modes.

\end{abstract}

\pacs{73.43.-f, 73.23.Hk, 73.23.-b}


\maketitle

Electron-electron interactions play an important role in an antidot in the integer quantum Hall regime~\cite{Sim08}. In an antidot with one or two bound modes (edge states), the number of which is determined by local filling factor $\nu_c$ around the antidot, the interactions cause interesting phenomena~\cite{Ford94,Maasilta,Kataoka99,Karakurt,Kataoka02,Sim03,Ihnatsenka,Kato},
such as charging effects and $h/2e$ Aharonov-Bohm (AB) effects. It is valuable to extend the phenomena to generic effects in antidots with multiple modes. The extension is reminiscent of the stream of studies from a quantum dot to multiple dots~\cite{Wiel02}, and useful for applying antidots to the fractional quantum Hall regime~\cite{Maasilta00,Averin07} or to qubit implementation~\cite{Averin01}.

\begin{figure}[b]
\centering\includegraphics[width=0.47\textwidth]{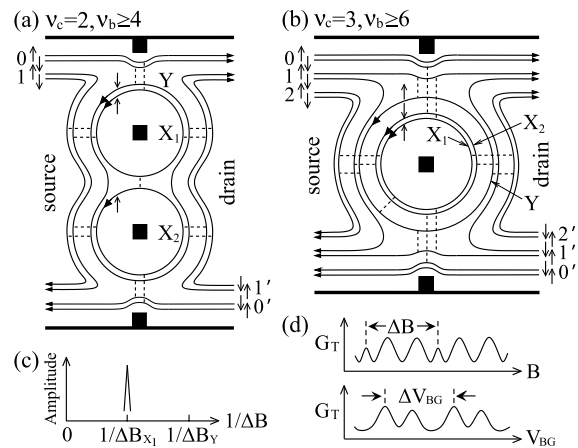}\\
\caption{Schematic views of (a-b) antidots and (c-d) relevant experimental data.
(a) A symmetric antidot molecule with local filling factor $\nu_c = 2$ and bulk filling $\nu_b \geq 4$. It has atomic modes X$_1$, X$_2$, and molecular mode Y.
Each mode has Landau-level (0, 1, 2) and spin ($\uparrow$, $\downarrow$) indexes. The solid and dashed lines represent edge states and electron tunneling, respectively.
(b) A $\nu_c = 3$ antidot with modes X$_1$, X$_2$, Y. 
(c) Fourier transformation of AB oscillations of conductance $G_T$ through an antidot molecule with $\nu_c = 4$. From~\cite{Gould96}.
(d) Magnetic-field $B$ and backgate-voltage $V_\textrm{BG}$ dependence of $G_T$ for an antidot with $\nu_c = 3$. From Figs. 11 and 13 of~\cite{Goldman08}.}
\label{AD-1-Setup}
\end{figure}

Some works~\cite{Gould96,Goldman08,Ihnatsenka09} have been done in that direction, but require further studies. In Ref.~\cite{Gould96}, an antidot molecule with $\nu_c=4$ was experimentally studied; see a simplified view with $\nu_c =2$ in Fig.~\ref{AD-1-Setup}(a). It has atomic modes $\mathrm{X_1}$, $\mathrm{X_2}$, and molecular modes $\mathrm{Y}$. Under certain conditions, AB resonances with period $\Delta B_\mathrm{Y}$ corresponding to the area enclosed by $\mathrm{Y}$ disappear in electron conductance $G_T$ through the system,
while those to $\mathrm{X_{1,2}}$ were observed [Fig.~\ref{AD-1-Setup}(c)].
This finding disagrees with the noninteracting electron case, in which $\mathrm{Y}$ more clearly shows AB resonances than $\mathrm{X_{1,2}}$ since Y couples more strongly with extended edge channels. Such disappearance of AB resonances was called spectator behavior~\cite{Gould96}. Its mechanism remains unclear despite of efforts~\cite{Takagaki97}.

Unexpected experimental results~\cite{Goldman08} were also found in a $\nu_c = 3$ antidot with three modes [Fig.~\ref{AD-1-Setup}(b)]. The magnetic-field dependence of $G_T$ shows three peaks in one AB period $\Delta B$, two of which have almost same peak height higher than the third [Fig.~\ref{AD-1-Setup}(d)].
And, the dependence of $G_T$ on the backgate voltage $V_\textrm{BG}$ applied to the antidot shows two alternating peak separations, i.e., the pairing of two neighboring peaks. Even more strange is that the peak pairing was not found for $\nu_c=2$ and 4. These features disagree with the noninteracting case, in which
there appear three independent peaks with different height and separation within one period since each mode shows one peak per period and couples to extended edge channels differently from the others; for example, $\mathrm{X_1}$ couples to channels $1\uparrow$ and $2'\uparrow$, while $\mathrm{X_2}$ to $0\downarrow$ and $1'\downarrow$ [Fig.~\ref{AD-1-Setup}(b)]. The unexpected results may come from the interactions, however, they are different from the $h/3e$ AB effect, a naive extension of the $h/2e$ AB effect~\cite{Ford94} of $\nu_c = 2$, in which the three peaks have the same height.

In this Letter, we theoretically study AB resonances in antidot systems with three modes in the integer quantum Hall regime, based on a capacitive interaction model. We predict the {\em spectator} behavior that the AB resonances of some modes disappear and instead are replaced by those of other modes, because of internal charge relaxation between the modes. Which and how many modes show the spectator behavior depends on ratios of capacitances. Our finding provides unified understanding of the unexpected results~\cite{Gould96,Goldman08} on the two different systems.

\emph{Antidots with three modes}.---
We consider two representative systems with three modes, a symmetric $\nu_c = 2$ molecule and a $\nu_c = 3$ antidot~\cite{Goldman08} (Fig.~\ref{AD-1-Setup}); a $\nu_c = 4$ molecule in Ref.~\cite{Gould96} is spin-unresolved so that some of its features can be described by the $\nu_c = 2 $ molecule. Each mode tunnel-couples to extended edge channels with the same spin, and also to the other modes with the same spin. The two systems have one outermost mode Y and two inner modes X$_1$ and X$_2$. We treat $\mathrm{X_1}$ and $\mathrm{X_2}$ equally, since in the molecule they are symmetric, and in the $\nu_c = 3$ antidot they have Zeeman gap (albeit exchange enhanced~\cite{Xu}) much smaller than Landau gap.

We consider the regime of zero bias, zero temperature, and strong perpendicular magnetic field $B_0 \gg \Delta B_\alpha$. Here, $\Delta B_\alpha$ is the AB period of mode $\alpha = \mathrm{X_1}, \mathrm{X_2}, \mathrm{Y}$. In the tunneling regime, we describe the total energy of the two systems by the same form (with system-dependent parameters) of the capacitive interaction model,
\begin{eqnarray}
E(\{\delta Q_{\alpha}\}) = \sum_{\alpha m} \tilde{\xi}_{\alpha m} n_{\alpha m} +
\sum_{\alpha \alpha'} U_{\alpha\alpha'} \delta Q_{\alpha} \delta Q_{\alpha'}/e^2, \label{TotalEnergy}
\end{eqnarray}
where $\alpha, \alpha' = \mathrm{X_1}, \mathrm{X_2}, \mathrm{Y}$. We derive it by generalizing the case of a $\nu_c=2$ antidot~\cite{Sim08,Sim03} in the same way as in multiple dots~\cite{Wiel02}. It governs the ground-state transition as a function of $B_0$ or $V_\textrm{BG}$. By analyzing the transition, we predict the features (height and separation) of AB resonance peaks in $G_T$. For the $\nu_c = 2$ case, the model~\eqref{TotalEnergy} successfully describes the charging effect, $h/(2e)$ AB effect and Kondo effect~\cite{Sim03}; we here ignore Kondo effects.

The first term of Eq.~\eqref{TotalEnergy} comes from the energy $\tilde{\xi}_{\alpha m}$ and occupation $n_{\alpha m}$ of single-electron
orbital $m$ of $\alpha$. We will derive below that $\tilde{\xi}$ satisfies
$\tilde{\xi}_{\alpha m} = \xi_{\alpha m, 0} + \Delta \xi_\alpha \delta B/\Delta B_\alpha$ when the magnetic field varies from $B_0$ by $\delta B$ $(\ll B_0)$.
Here $\Delta  \xi_\alpha$ is the single-particle level spacing of $\alpha$,
and $\tilde{\xi}_{\alpha m} = \xi_{\alpha m,0}$ at $B_0$. This dependence of $\tilde{\xi}_{\alpha m}$ on $\delta B$ leads to the fact that mode $\alpha$ shows one AB resonance per period $\Delta B_\alpha$ in the noninteracting limit.

The second term of Eq.~\eqref{TotalEnergy} shows capacitive interactions $U_{\alpha \alpha'} \equiv e^2 (C^{-1})_{\alpha \alpha'} / 2$ between the excess charges $\delta Q_\alpha$ accumulated in mode $\alpha$. $\delta Q_\alpha$ depends on $\delta B$ as
\begin{equation}\label{ExcessCharge}
\delta Q_{\alpha} = e N_{\alpha} - Q_{\alpha}^\mathrm{G} + e\,\delta B/\Delta B_\alpha.
\end{equation}
The total charge $eN_{\alpha}$ of $\alpha$ is compensated by gate charge $Q_\alpha^\mathrm{G}$ ($\propto V_\textrm{BG}$) tuned by $V_\textrm{BG}$. $N_\alpha (= \sum_m n_{\alpha m})$ varies by an integer due to the discreteness of electron charge $e < 0$.

We explain the dependence of $\tilde{\xi}_{\alpha m}$ and $\delta Q_\alpha$ on $\delta B$. As $B_0$ increases by $\delta B$, each orbital $\alpha m$ spatially shifts toward the center of its antidot to keep enclosing a given number, saying $m$, of magnetic flux quanta. Then, its energy changes by $(\Delta \xi_\alpha + 2\sum_{\alpha'} U_{\alpha' \alpha } \delta Q_{\alpha'}/eN_\alpha) \delta B / \Delta B_\alpha$. The term $\Delta \xi_\alpha \delta B / \Delta B_\alpha$, coming from antidot potential, results in the dependence of $\tilde{\xi}_{\alpha m}$ on $\delta B$. The other term, resulting from the interactions between the orbital and $\delta Q_{\alpha'}$, causes the dependence on $\delta B$ in Eq.~\eqref{ExcessCharge}. The dependence on $\delta B$ captures the physics of antidots.

We discuss the parameters of Eq.~\eqref{TotalEnergy}. For $B_0 \gg \Delta B_\alpha$, it is natural to apply the constant interaction model~\cite{Wiel02} that $\Delta \xi_\alpha$ and $C_{\alpha \alpha'}$ are constant over several AB periods, and that $C_{\alpha\alpha} = |C_{g, \alpha}| + \sum_{\alpha' \ne \alpha} |C_{\alpha \alpha'}|$. $C_{g, \alpha}$ is the ``gate'' capacitance of $\alpha$ due to extended edge channels and $V_\textrm{BG}$. $\mathrm{X_1}$ and $\mathrm{X_2}$ have the same values of $\Delta B_\alpha$, $\Delta \xi_\alpha$, $U_{\alpha \alpha}$, $U_{\alpha \mathrm{Y}}$, and $C_{g, \alpha}$ because of the symmetry.

\emph{Charge accumulation and relaxation}.---
As $\delta B$ increases, $\delta Q_\alpha$ continuously accumulates with rate $1/\Delta B_\alpha$ as in Eq.~\eqref{ExcessCharge}. The accumulated charges are relaxed with resonant tunneling, resulting in the transition of the ground-state configuration $(N_\mathrm{X_1}, N_\mathrm{X_2}, N_\mathrm{Y})$. There are two kinds of single-electron relaxation. External relaxation occurs between a mode (here, $\mathrm{X_1}$) and extended edge channels (with Fermi level $\epsilon_F$), e.g., when $E(\delta Q_{\mathrm{X_1}} \pm e,\delta Q_{\mathrm{X_2}},\delta Q_{\mathrm{Y}}) = E(\delta Q_{\mathrm{X_1}},\delta Q_{\mathrm{X_2}},\delta Q_{\mathrm{Y}}) \pm \epsilon_F$. This causes resonance peaks in $G_T$. By contrast, {\em internal} relaxation occurs between modes, through tunneling or cotunneling mediated by virtual states, e.g., when
\begin{equation}\label{CotunCon}
E(\delta Q_{\mathrm{X_1}} \pm e,\delta Q_{\mathrm{X_2}},\delta Q_{\mathrm{Y}} \mp e) = E(\delta Q_{\mathrm{X_1}},\delta Q_{\mathrm{X_2}},\delta Q_{\mathrm{Y}}).
\end{equation}
It does not cause peaks in $G_T$. It occurs only between Y and $\alpha$ $\in \{ \mathrm{X_1}, \mathrm{X_2} \}$ in our case with the symmetry between $\mathrm{X_1}$ and $\mathrm{X_2}$. In general, relaxations involving more than one electron can occur. Two-electron relaxation occurs in the molecule (see below), while not in the $\nu_c=3$ antidot.

\begin{figure}[t]
\centering\includegraphics[width=0.47\textwidth]{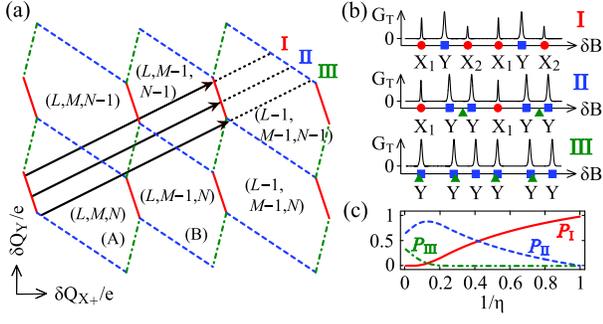}\\
\caption{(color online) (a) Charge stability diagram for a $\nu_c = 3$ antidot. It consists of two types (A) and (B) of hexagonal cells in $(\delta Q_\mathrm{X_+},\delta Q_\mathrm{Y})$ plane. Each cell represents a ground-state configuration of $(N_{\mathrm{X_1}},N_{\mathrm{X_2}},N_{\mathrm{Y}}) = (L,M,N)$.
Cell boundaries are determined by charge relaxation conditions such as Eq.~\eqref{CotunCon}. At dashed blue (solid red) boundaries, AB resonances occur via tunneling through Y ($\mathrm{X_1}$ or $\mathrm{X_2}$), and at dash-dot green boundaries internal charge relaxations occur between Y and $\mathrm{X_{1,2}}$.
As the magnetic field increases, $(\delta Q_\mathrm{X_+},\delta Q_\mathrm{Y})$ evolve along a line (solid arrow) of slope $\Delta B_{\mathrm{X_1}}/ (2\Delta B_{\mathrm{Y}})$, while $\delta Q_\mathrm{X_-}$ is constant within a cell and
differs by charge $e$ between (A) and (B). Depending on initial values of $\delta Q_\alpha$'s at given field $B_0$, the evolution shows one of three possible sequences of AB resonances, ``$\mathrm{X_1}$-Y-$\mathrm{X_2}$" (type I), ``$\mathrm{X_1}$-Y-Y" (type II), ``Y-Y-Y" (type III). Parameters are chosen as
$\Delta \xi_\alpha = 0$, $\Delta B_\mathrm{X_1} = \Delta B_\mathrm{Y}$, $C_{g,\mathrm{Y}} = 8 C_{g,\mathrm{X_1}}$, $C_{\mathrm{X_1} \mathrm{Y}} = 2 C_{g, \mathrm{X_1}}$, $C_{\mathrm{X_1} \mathrm{X_2}} = 8 C_{\mathrm{X_1} \mathrm{Y}}$,
and $\delta Q_\mathrm{X_-} = 0$ for cell (A) and $- e$ for (B).
(b) Sequence of resonance peaks in $G_T$ as a function of $\delta B$ for each type shown in (a). The modes giving peaks are shown. Triangles represent internal charge relaxation.
(c) Probability $P_J (1/\eta)$ of finding the sequences of type $J = \mathrm{I, II, III}$ is drawn with the parameters of (a).}
\label{AD-2-Evolution}
\end{figure}

The ground-state evolution of the antidots and the resulting AB resonances are governed by the relaxations. We study them by analyzing a charge stability diagram~\cite{Wiel02}. In Fig.~\ref{AD-2-Evolution}(a), it is drawn for a $\nu_c = 3$ antidot in $(\delta Q_{\mathrm{X_+}},\delta Q_{\mathrm{Y}})$ plane, where $\delta Q_{\mathrm{X_{\pm}}} \equiv \delta Q_{\mathrm{X_1}} \pm \delta Q_{\mathrm{X_2}}$; this two-dimensional view is possible due to the symmetry of
$\mathrm{X_1}$ and $\mathrm{X_2}$. Below, we first consider the strong interaction regime of $U_{\alpha \alpha} \gg \Delta\xi_\alpha$, which is analogous to the case of metallic dots, and then discuss finite $\Delta \xi_\alpha$.

The internal relaxation results in the spectator behavior. The evolution of $\{ \delta Q_\alpha \}$ follows different types of sequences of AB resonances, depending on how many times the spectator behavior appears per $\Delta B_\mathrm{X_1}$. For example, in a $\nu_c=3$ antidot, there are three types I, II, III [Fig.~\ref{AD-2-Evolution}]. In type I of X$_1$-Y-X$_2$, the evolution never passes the internal relaxation, and AB resonances occur sequentially by X$_1$, Y, X$_2$, X$_1$, Y, X$_2$, $\cdots$. In type II of X$_1$-Y-Y (III of Y-Y-Y), the evolution passes the internal relaxation once (twice) per $\Delta B_\mathrm{X_1}$, and the AB resonances by X$_2$ (X$_1$ and X$_2$) disappear
and are replaced by those by Y. Here, X$_1$ or X$_2$ shows the spectator behavior.

We discuss the general features of the spectator behavior. Which mode shows the behavior is governed by
\begin{equation}
\eta \equiv \frac{(U_{\mathrm{X_1 X_1}} + U_{\mathrm{X_1 X_2}} - 2 U_{\mathrm{X_1 Y}})\delta B / \Delta B_\mathrm{X_1}}{(U_{\mathrm{YY}} - U_{\mathrm{X_1 Y}})\delta B / \Delta B_\mathrm{Y}} = \frac{C_{g,\mathrm{Y}} \Delta B_\mathrm{Y}}{C_{g,\mathrm{X_1}} \Delta B_\mathrm{X_1}},
\label{C_ratio}
\end{equation}
the ratio of energy gains between $\delta Q_\mathrm{X_+}$ $(= \delta Q_\mathrm{X_1} + \delta Q_\mathrm{X_2})$ and $\delta Q_\mathrm{Y}$ in the internal relaxation between them; see Eq.~\eqref{CotunCon}. In Fig.~\ref{AD-2-Evolution}(a), $\eta$ equals the ratio of slopes between the dash-dot relaxation line and the evolution arrow. When $\eta < 1$, the relaxation occurs from $\delta Q_\mathrm{Y}$ to $\delta Q_\mathrm{X_1}$ or $\delta Q_\mathrm{X_2}$; hence Y shows the spectator behavior. For $\eta > 1$, X$_1$ and X$_2$ show it. On the other hand, the inter-mode interaction strength ($ \propto |C_{\alpha' \alpha (\ne \alpha')}|$) also governs the behavior. As the strength increases and as $\eta$ more and more deviates from 1, more sequences (with different ``initial'' values of $\{ \delta Q_\alpha \}$ at $B_0$) show the behavior (type II or III); the dash-dot line in Fig.~\ref{AD-2-Evolution}(a) becomes longer so that the evolution has more chance to pass the internal relaxation. When the strength vanishes or $\eta = 1$, the spectator behavior is suppressed and only type I appears. Note that noninteracting electrons show only type I.

\emph{Antidot with $\nu_c=3$}.---
We discuss the spectator behavior in a $\nu_c = 3$ antidot [Fig.~\ref{AD-1-Setup}(b)]. Its geometry indicates $\Delta B_\mathrm{X_1} \simeq \Delta B_\mathrm{Y}$ and $|C_{\mathrm{X_1X_2}}| > |C_{\mathrm{X_1Y}}|$. The spatial separation between the outermost mode Y (inner modes $\mathrm{X_{1,2}}$)
and the extended channels $1\downarrow$ is governed by Zeeman splitting energy
(Landau gap) so that $C_{g, \mathrm{Y}} / C_{g, \mathrm{X_1}}$ is much larger than 1. These lead to $\eta > 1$ [Eq.~\eqref{C_ratio}]. Hence, as mentioned above, $\mathrm{X_1}$ and $\mathrm{X_2}$ show the spectator behavior that type II sequence of $\mathrm{X_1}$-Y-Y and III of Y-Y-Y appear instead of I of $\mathrm{X_1}$-Y-$\mathrm{X_2}$. As $\eta$ ($> 1$) and $|C_\mathrm{X_{1,2} Y}|$ increase, type II and III appear more dominantly. For a $\nu_c =3$ antidot, we obtain the probability $P_J (1/\eta)$ of finding type $J \in \{ \mathrm{I},
\mathrm{II}, \mathrm{III}  \}$ in the ensemble of sequences with different initial values of $\{ \delta Q_\alpha \}$ at $B_0$ [Fig.~\ref{AD-2-Evolution}(c)].

In Fig.~\ref{AD-2-Evolution}(b), we plot $G_T (\delta B)$. We obtain it
in the sequential tunneling regime using the standard master equation method~\cite{Ihnatsenka09}, which is enough for demonstrating the positions and relative heights of AB peaks; here, we assumed low temperature ($\ll U_{\alpha \alpha}$) and the backward-reflection regime, as in Ref.~\cite{Goldman08},
that mode $\alpha$ couples to edge channel $\beta$ with coupling strength $\gamma_{\alpha-\beta}$ as $\gamma_\mathrm{Y-1\uparrow} > \gamma_\mathrm{X_1 - 1\uparrow} > \cdots$. For each type, we describe the features of $G_T$. In type I, each of X$_1$, X$_2$, and Y shows one peak per period $\Delta B_\mathrm{X_1}$. The resulting three peaks in $\Delta B_\mathrm{X_1}$ have different height, because of different $\gamma_{\alpha-\beta}$'s. The peak by Y is the highest, since Y is the outermost mode.

In type II, two peaks among the three within $\Delta B_\mathrm{X_1}$ come from Y, and have the same height higher than the third. The separation $\kappa \Delta B_\mathrm{X_1}$ between two consecutive peaks by Y depends on interactions as
$\kappa = U_{\mathrm{X_1Y}}/(2U_{\mathrm{X_1Y}} + U_{\mathrm{YY}})$, regardless of the initial values of $\{ \delta Q_\alpha  \}$ at $B_0$. In the strong inter-mode interaction limit of $C_{g, \mathrm{X_1}}/C_{\mathrm{X_1Y}} \to 0$,
$\kappa \to 1/3$. The position of the other peak by $\mathrm{X_1}$ or $\mathrm{X_2}$ depends on the initial values of $\{ \delta Q_\alpha \}$.

In type III, all the three peaks within $\Delta B_\mathrm{X_1}$ come from Y,
showing the same peaks. The separation between them is determined by interactions as $\kappa \Delta B_\mathrm{X_1}$ and $(1 - 2 \kappa) \Delta B_\mathrm{X_1}$.
In the strong inter-mode interaction limit, it becomes $\Delta B_\mathrm{X_1} / 3$, showing $h/(3e)$ AB effects, and the total energy in Eq.~\eqref{TotalEnergy}
becomes  $E \simeq U \delta Q_\textrm{tot}^2 / e^2$, where $\delta Q_\textrm{tot} = \sum_\alpha \delta Q_\alpha = 3 e \delta B / \Delta B_\mathrm{X_1} + \cdots$.
This form of $E$, mentioned in literatures~\cite{Ihnatsenka09}, cannot describe type II.

So far, we have restricted to $\Delta \xi_\alpha = 0$. In the case of finite level spacing $\Delta \xi_\alpha$, the first term of Eq.~\eqref{TotalEnergy} is absorbed into the second so that $E$ has the same form as that of $\Delta \xi_\alpha = 0$, but with replacement (i) $U_{\alpha \alpha} \to U_{\alpha \alpha} + \Delta \xi_\alpha / 2$ and (ii) $Q_\alpha^{\mathrm{G}} \to \tilde{Q}_\alpha^{\mathrm{G}}$, where $\tilde{Q}_\alpha^{\mathrm{G}}$ is obtained by $\sum_{\alpha'} (U_{\alpha \alpha'} + \delta_{\alpha \alpha'} \Delta \xi_\alpha/2 ) \tilde{Q}_{\alpha'}^{\mathrm{G}} = \sum_{\alpha'} U_{\alpha \alpha'} Q_{\alpha'}^{\mathrm{G}}$. The replacement does not modify the dependence of $\delta Q_\alpha$ on $\delta B$ in Eq.~\eqref{ExcessCharge}, but weakens the spectator behavior [see replacement (i)]. For example, when $\Delta \xi_\alpha$ is comparable to $e^2 / C_\mathrm{X_1 X_1}$, type II and III are suppressed by 25~\% and 100~\%, respectively, for the antidot studied in Fig.~\ref{AD-2-Evolution}.

The above findings indicate that the result of Ref.~\cite{Goldman08}, two peaks with equal height in $\Delta B$ [the upper panel of Fig.~\ref{AD-1-Setup}(d)], may be explained by type II; we do not exclude the possibility of type I that
two modes among the three accidently give the two peaks with almost equal height.

On the other hand, replacement (ii) affects the evolution of $\delta Q_\alpha$ as a function of $V_\mathrm{BG}$. For $\Delta \xi_\alpha = 0$, the evolution follows a line of slope $Q_\mathrm{Y}^\mathrm{G} / (2 Q_\mathrm{X_1}^\mathrm{G}) \simeq 0.5$ in the stability diagram. When $\Delta \xi_\alpha$ is finite, the slope becomes $s = \tilde{Q}_\mathrm{Y}^\mathrm{G} / (2\tilde{Q}_\mathrm{X_1}^\mathrm{G}) \simeq 0.5[1 + (|C_{g, \mathrm{X_1}}| + 3 |C_\mathrm{X_1 Y}|) \Delta \xi_\mathrm{X_1} / e^2]/ [1 + (|C_{g, \mathrm{Y}}| + 3 |C_\mathrm{X_1 Y}|) \Delta \xi_\mathrm{Y} / e^2]$. $s$ can be very small for $\Delta \xi_\alpha \simeq e^2 / |C_\mathrm{X_1 Y}|$, $|C_{g, \mathrm{Y}}| \gg |C_{g, \mathrm{X_1}}|$, $|C_\mathrm{X_1 Y}|$. The latter condition can be satisfied in a $\nu_c = 3$ antidot since the spatial separation between Y (X$_1$) and extended edge channels is determined by Zeeman (Landau) splitting. The evolution line with small slope $s$ can pass only the solid boundaries
in the stability diagram [Fig.~\ref{AD-2-Evolution}], showing paired peaks by X$_1$ and X$_2$. Or, depending on the initial value of $\{ Q_\alpha \}$
at $B_0$, it can pass only the dashed lines, showing paired peaks by Y.
The paired peaks agree with the lower panel of Fig.~\ref{AD-1-Setup}(d). In the cases of $\nu_c = 2$ and 4 with finite $\Delta \xi_\alpha$, the slope $s$ has a similar form to $\nu_c = 3$, but the peak pairing does not appear, because
the separation between the outermost mode and edge channels is governed by Landau splitting so that $s$ has a similar value to the case of $\Delta \xi_\alpha=0$. These indicate that the peak pairing in Ref.~\cite{Goldman08} is due to finite $\Delta \xi_\alpha$ and the interaction between Y and edge channels.

\emph{Molecule}.---
We discuss the molecule in Fig.~\ref{AD-1-Setup}(a). Its geometry implies $\Delta B_\mathrm{X_1} \gtrsim 2 \Delta B_\mathrm{Y}$ and $|C_{\mathrm{X_1X_2}}| < |C_\mathrm{X_{1} Y}|$. Since the circumference of Y is shorter than two times of that of $\mathrm{X_{1}}$, one has $|C_{g,\mathrm{Y}}| < 2  |C_{g, \mathrm{X_1}}|$, provided that $V_\mathrm{BG}$ affects  $C_{g,\mathrm{Y}}$ and $C_{g, \mathrm{X_1}}$ more dominantly than edge channels. Then, $\eta < 1$, and Y shows the spectator behavior.

\begin{figure}[t]
\centering\includegraphics[width=0.46\textwidth]{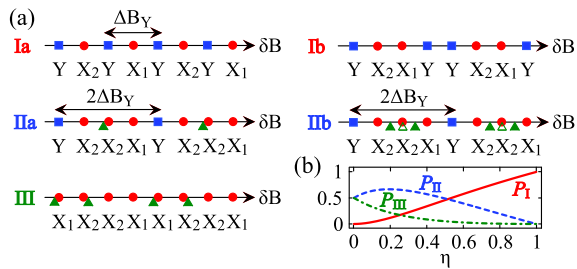}\\
\caption{(color online) (a) Selected types of sequences of AB resonances, as a function of $\delta B$ for a $\nu_c = 2$ molecule. Filled and empty triangles
represent the single- and two-electron internal relaxations, respectively.
(b) Probability $P_J (\eta)$ of finding the sequences of type $J = \mathrm{I, II, III}$. Here, I (II) means the types, e.g., Ia and Ib (IIa and IIb), having two (one) Y-resonances within $\Delta B_\mathrm{X_1}$. Parameters are chosen as $\Delta \xi_\alpha = 0$, $\Delta B_\mathrm{X_1} = 2 \Delta B_\mathrm{Y}$,
$C_{\mathrm{X_1} \mathrm{Y}} = 10 C_{g, \mathrm{X_1}}$, and $C_{\mathrm{X_1} \mathrm{X_2}} = 0.5 C_{\mathrm{X_1} \mathrm{Y}}$.}
\label{AD-3-ABPatterns}
\end{figure}

For an example case of $C_\mathrm{X_1 X_2} = 0.5 C_\mathrm{X_1 Y}$, selected AB resonance sequences are shown as a function of $\delta B$ in Fig.~\ref{AD-3-ABPatterns}. The spectator behavior occurs such that type Ia of Y-X$_2$-Y-X$_1$ is replaced by IIa of Y-X$_2$-X$_2$-X$_1$ (III of X$_1$-X$_2$-X$_2$-X$_1$) when the relaxation in Eq.~\eqref{CotunCon} occurs once (twice) within $\Delta B_\mathrm{X_1}$. In the molecule, in addition to the one-electron relaxation, there occurs two-electron relaxation, $E(\delta Q_\mathrm{X_1} \pm e, \delta Q_\mathrm{X_2} \pm e, \delta Q_\mathrm{Y} \mp e) = E(\delta Q_\mathrm{X_1}, \delta Q_\mathrm{X_2}, \delta Q_\mathrm{Y}) \pm \epsilon_F$, which is a mixture of internal and external relaxations. This additional process results in more types such as IIb of Y-X$_2$-X$_2$-X$_1$.
Type IIb has the same sequence as IIa, and results from Ib of Y-X$_2$-X$_1$-Y
due to two one-electron internal relaxations and  one two-electron relaxation within $\Delta B_\mathrm{X_1}$. We plot $P_J (\eta)$ of type $J$ in Fig.~\ref{AD-3-ABPatterns}(b). As $\eta$ decreases from 1, type II (e.g., IIa and IIb) becomes dominant. Note that when $\Delta \xi_\alpha \simeq e^2 / C_\mathrm{\alpha \alpha}$, $P_\mathrm{II}$ and $P_\mathrm{III}$ are reduced by less than 10~\% for $0.6 \lesssim \eta < 1$. Qualitatively same features appear in other parameter ranges.

The unexpected results of Ref.~\cite{Gould96} can be understood by type II, provided that the inner modes X$_1$ and X$_2$ are almost decoupled from edge channels (i.e., only Y shows AB peaks). In this case, the period of AB peaks in type II is $2 \Delta B_\mathrm{Y}$ instead of $\Delta B_\mathrm{Y}$ due to the spectator behavior [Fig.~\ref{AD-3-ABPatterns}(a)]. This agrees with Fig.~\ref{AD-1-Setup}(c), since $2 \Delta B_\mathrm{Y} \simeq \Delta B_\mathrm{X_1}$. On the contrary, all the other types of $\eta < 1$ and those of $\eta > 1$ cannot explain Fig.~\ref{AD-1-Setup}(c); for example, type I shows
peaks with $\Delta B_\mathrm{Y}$ or a mixture of $\Delta B_\mathrm{Y}$ and $\Delta B_\mathrm{X_{1}}$, depending on the coupling of X$_{1,2}$ with edge channels. These indicate that the molecule of Ref.~\cite{Gould96} is in the regime of type II of $\eta < 1$.

\emph{Conclusion}.---
Electron-electron interactions give rise to the spectator behavior of AB resonances in antidots with three modes. The spectator behavior is generic, i.e., expected to appear in other quantum Hall systems with multiple modes, such as antidots and quantum dots. And it is useful for detecting interactions between edge states. To experimentally test the spectator behavior and our explanation of the experimental data~\cite{Gould96,Goldman08}, one can monitor the modes showing resonance signals by selective injection and detection of edge channels~\cite{Kataoka03}.

We thank C. J. B. Ford, V. J. Goldman, and M. Kataoka for discussion, and NRF (2009-0078437).


\begin{thebibliography}{99}

\bibitem{Sim08} H.-S. Sim, M. Kataoka, and C. J. B. Ford, Phys. Rep. {\bf 456}, 127 (2008).
\bibitem{Ford94} C. J. B. Ford {\it et al.},
    Phys. Rev. B {\bf 49}, 17456 (1994);
    M. Kataoka {\it et al.},
    Phys. Rev. B {\bf 62}, R4817 (2000).
\bibitem{Maasilta} I. J. Maasilta and V. J. Goldman, Phys. Rev. B {\bf 57} R4273 (1998).
\bibitem{Kataoka99} M. Kataoka {\it et al.},
    Phys. Rev. Lett. {\bf 83}, 160 (1999).
\bibitem{Karakurt} I. Karakurt {\it et al.},
    Phys. Rev. Lett. {\bf 87} 146801 (2001).
\bibitem{Kataoka02} M. Kataoka, C. J. B. Ford, M. Y. Simmons, and D. A. Ritchie, Phys. Rev. Lett. {\bf 89}, 226803 (2002).
\bibitem{Sim03} H.-S. Sim {\it et al.},
    Phys. Rev. Lett. {\bf 91}, 266801 (2003);
    N. Y. Hwang, S.-R. E. Yang, H.-S. Sim, and H. Yi, Phys. Rev. B {\bf 70}, 085322 (2004).
\bibitem{Ihnatsenka} S. Ihnatsenka and I. V. Zozoulenko, Phys. Rev. B {\bf 74} 201303(R) (2006).
\bibitem{Kato} M. Kato {\it et al.},
    Phys. Rev. Lett. {\bf 102}, 086802 (2009).
\bibitem{Wiel02} W. G. van der Wiel {\it et al.},
    Rev. Mod. Phys. {\bf 75}, 1 (2002).
\bibitem{Maasilta00} I. J. Maasilta and V. J. Goldman, Phys. Rev. Lett. {\bf 84} 1776 (2000).
\bibitem{Averin07} D. V. Averin and J. A. Nesteroff, Phys. Rev. Lett. {\bf 99} 096801 (2007).
\bibitem{Averin01} D. V. Averin and V. J. Goldman, Solid State Commun. {\bf 121} 25 (2001).
\bibitem{Gould96} C. Gould {\it et al.},
    Phys. Rev. Lett. {\bf 77}, 5272 (1996).
\bibitem{Goldman08} V. J. Goldman, J. Liu, and A. Zaslavsky, Phys. Rev. B {\bf 77}, 115328 (2008).
\bibitem{Ihnatsenka09} S. Ihnatsenka, I. V. Zozoulenko, and G. Kirczenow, Phys. Rev. B {\bf 80}, 115303 (2009).
\bibitem{Takagaki97} Y. Takagaki, Phys. Rev. B {\bf 55}, R16021 (1997).
\bibitem{Xu} W. Xu {\it et al.}, J. Phys.: Condens. Matter {\bf 7}, 4419 (1995).
\bibitem{Kataoka03} M. Kataoka, C. J. B. Ford, M. Y. Simmons, and D. A. Ritchie, Phys. Rev. B {\bf 68} 153305 (2003).
    
\end{thebibliography}
\end{document}